\documentclass[fleqn,usenatbib]{mnras}
\usepackage{newtxtext,newtxmath}
\usepackage{color,soul}
\usepackage[T1]{fontenc}
\usepackage{ae,aecompl}
\usepackage{bm}

\usepackage{graphicx}	
\usepackage{amsmath}	
\usepackage{amssymb}	
\usepackage{todonotes}
\usepackage{multirow}

\title[Layering-related linear features on comet 67P]{Analysis of layering-related linear features on comet 67P/Churyumov-Gerasimenko}

\author[Ruzicka et al.]{
Birko-Katarina Ruzicka,$^{1,3}$\thanks{E-mail: ruzicka@mps.mpg.de}
Luca Penasa,$^{2}$
Hermann Boehnhardt,$^{1}$
Andreas Pack,$^{3}$
\newauthor Benoit Dolives,$^{4}$
Fabrice Souvannavong,$^{4}$
and Emile Remetean$^{5}$
\\
$^{1}$Max-Planck-Institut f\"ur Sonnensystemforschung, Justus-von-Liebig-Weg 3, 37077 G\"ottingen, Germany\\
$^{2}$Center of Studies and Activities for Space, CISAS, `G. Colombo', University of Padova, Via Venezia 15, 35131 Padova, Italy\\
$^{3}$Georg-August-Universit\"at G\"ottingen, Geowissenschaftliches Zentrum, Abteilung Isotopengeologie, Goldschmidtstra\ss e 1, 37073 G\"ottingen, Germany\\
$^{4}$Magellium, Imaging and Applications Department, 24 Rue Herm\`es, BP 12113, 31521 Ramonville Saint-Agne Cedex, France\\
$^{5}$Engineering for Future Missions Department, Centre National d'\'Etudes Spatiales, Bpi 1712, 18 av. E. Belin, 31401 Toulouse Cedex 9, France
}

\date{This article has been accepted for publication in MNRAS, Published by Oxford University Press, with the DOI 10.1093/mnras/sty3079}
\pubyear{2018}

\begin{document}
\label{firstpage}
\pagerange{\pageref{firstpage}--\pageref{lastpage}}
\maketitle

\begin{abstract}
We analysed layering-related linear features on the surface of comet 67P/Churyumov-Gerasimenko (67P) to determine the internal configuration of the layerings within the nucleus. We used high-resolution images from the OSIRIS Narrow Angle Camera onboard the Rosetta spacecraft, projected onto the SHAP7 shape model of the nucleus, to map 171 layering-related linear features which we believe to represent terrace margins and strata heads. From these curved lineaments, extending laterally to up to 1925\, m, we extrapolated the subsurface layering planes and their normals. We furthermore fitted the lineaments with concentric ellipsoidal shells, which we compared to the established shell model based on planar terrace features. Our analysis confirms that the layerings on the comet's two lobes are independent from each other. Our data is not compatible with 67P's lobes representing fragments of a much larger layered body. The geometry we determined for the layerings on both lobes supports a concentrically layered, `onion-shell' inner structure of the nucleus. For the big lobe, our results are in close agreement with the established model of a largely undisturbed, regular, concentric inner structure following a generally ellipsoidal configuration. For the small lobe, the parameters of our ellipsoidal shells differ significantly from the established model, suggesting that the internal structure of the small lobe cannot be unambiguously modelled by regular, concentric ellipsoids and could have suffered deformational or evolutional influences. A more complex model is required to represent the actual geometry of the layerings in the small lobe.
\end{abstract} 




\section{Introduction}

In-situ images of the nucleus surfaces of Jupiter-family comets (e.g., 9P, 81P, 103P) have previously been used to speculate about a layered structure of these nuclei \citep[e.g.,][]{thomas_shape_2007,bruck_syal_geologic_2013,cheng_surface_2013}. \citet{belton_internal_2007} proposed a model in which their interior consists of a core overlain by layerings that were locally and randomly piled onto the core through collisions (the `talps' or `layered pile' model). More recently, \citet{belton_origin_2018} suggested that the layerings could have been formed by fronts of self-sustaining amorphous to crystalline ice phase-change propagating from the nucleus surface to the interior. 

The ongoing debate about the origin of layerings in cometary nuclei might benefit from a more comprehensive understanding of their geometry and orientation. This understanding was greatly improved through the data collected by ESA's Rosetta mission to comet 67P/Churyumov-Gerasimenko (67P). Images taken by the OSIRIS camera system, surpassing the spatial resolution of previous missions by more than an order of magnitude, resolved features exposed across most of the nucleus surface which we interpret as layerings. Repetitive staircase patterns are formed by laterally persistent cliffs separating planar `terrace' surfaces. The cliff faces display parallel linear grooves, which are reminiscent of sedimentary outcrops on Earth where differential erosion carves such grooves into layerings of alternating hardness. The dust-free walls of deep pits reveal quasi-parallel sets of lineaments, oriented roughly perpendicular to the local gravity vector and extending to depths of at least a hundred meters below the present-day nucleus surface \citep{vincent_large_2015}. \\

A first systematic analysis of the orientation of 67P's layerings was conducted by \citet{massironi_two_2015}. They created a series of geologic cross-sections of the comet nucleus based on the orientation of planes fitted to morphologically flat areas (`terraces') on a shape model of the nucleus surface. \citet{massironi_two_2015} concluded that the two lobes of the nucleus are independently-formed bodies with an `onion-shell' layered inner structure that formed before the two bodies merged in a gentle collision to form the nucleus of 67P. 

Using a similar approach, albeit on a much smaller number of measurements, \citet{rickman_comet_2015} suggested that morphological ridges and other features on opposing sides of the comet's lobes can be connected by planar features. They interpreted this correlation as evidence for a semi-planar, pervasive internal layering. This would suggest that the two lobes of 67P are fragments of a much larger body.

\citet{penasa_three_2017} modeled the inner layered structure of comet 67P by fitting concentric ellipsoidal shells to a total of 483 terraces on both lobes, providing a first simplified 3D geological model of their inner structure. By comparing the orientation of the surface planes with the two model ellipsoids, they suggest that the inner structure of the two lobes can be explained by a set of concentric ellipsoids.

\begin{figure}
	\includegraphics[width=\columnwidth]{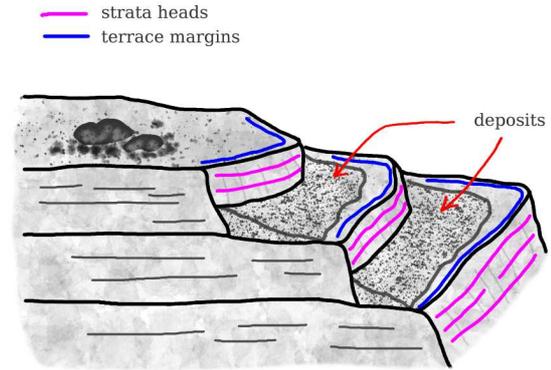}
    \caption{Schematic illustration of the two types of layering-related linear features we analyzed in this paper: i) morphological terrace margins (red curves) and ii) strata heads visible as lineaments on cliff faces and hill slopes (blue curves). Bedding orientations derived from these linear features are not affected by deposits on the terraces.}
    \label{fig:figure1}
\end{figure}

The studies by \citet{massironi_two_2015}, \citet{rickman_comet_2015}, and \citet{penasa_three_2017} made use of terraces as a proxy for the underlying orientation of the layers. The advantage of such an approach is that terraces are ubiquitous on the cometary body, that they can easily be mapped, and that their orientation can be easily estimated by means of a best fitting procedure applied to the vertices of a shape model \citep{massironi_two_2015}. A downside of the terrace approach is that results may be biased due to airfall and mass wasting processes from nearby cliffs (\autoref{fig:figure1}). \citet{penasa_three_2017} acknowledge this limitation and estimate that this might introduce an error of up to $\sim$20\textdegree{} to their normals. 

Here we use layering-related linear features instead of terraces to avoid possible bias due to depositional processes. We studied two types of linear features (\autoref{fig:figure1}): 

i) morphological edges along terrace margins, adjacent to cliffs (referred to as `terrace margins') and 

ii) lineaments on hill slopes and cliff faces \citep[referred to as `strata heads', in accordance with][]{lee_geomorphological_2016} produced by the intersection of the layers with the topographic surface. Both types of features appear to be erosional consequences of discontinuities between the individual layerings, which are possibly related to planes of different physical properties within layered materials. The linear traces mark the locations where the inner bedding planes intersect with the topography at a high angle \citep{massironi_two_2015}. 

\section{Data and Methods}

\subsection*{Data base}

For the three-dimensional representation of the nucleus of 67P we used the SHAP7 shape model, which is obtained from a stereo-photogrammetric (SPG) analysis of images taken by the OSIRIS Narrow Angle Camera \citep[NAC;][]{keller_osiris_2007} onboard Rosetta. The model covers the whole nucleus and consists of about 44 million facets, has a mean accuracy of 0.3\,m at a horizontal sampling of about 1-1.5\,m, and vertical accuracy at the decimeter scale \citep{preusker_global_2017}. 

Most of our mapping was conducted on high-resolution images of the nucleus surface. We selected suitable OSIRIS NAC images, publicly available and retrieved from the ESA Planetary Science Archive (PSA), according to these criteria: Images taken with the OSIRIS NAC orange filter F22 (which have a high signal-to-noise ratio; images taken at a spacecraft-comet distance of <\,30\,km (resulting in a pixel resolution of 0.5 to 1.5\,m at the image centre); images where the illumination is both sufficiently bright and also in a suitable direction to show layering-related features in good contrast. Initially, we selected only images that were calibrated for geometrical distortion due to the internal camera geometry \citep[CODMAC level 4;][]{tubiana_scientific_2015}. We later decided to supplement these with geometrically uncalibrated images (level 3) in order to improve coverage of the nucleus. We minimized the impact of this distortion on our data by restricting our mapping efforts to the central area of uncalibrated images, where the distortion decreases to near-zero.

\subsection*{Mapping of linear features}

Large-scale morphological terrace margins were mapped directly on the shape model by manually tracing the features as polylines in \textsc{CloudCompare} \citep{girardeau-montaut_2014_cloudcompare}.  

Mapping the finer morphological details of strata heads required a higher resolved mapping medium, for which we projected two-dimensional OSIRIS images onto the three-dimensional shape model. We used a customized version of the software \textsc{philae localization workshop} \citep[PLW,][]{remetean_philae_2016} for the measurement process of the linear features. For each image, we manually selected a set of reference points, consisting of prominent landmarks visible on both the image and the shape model (e.g., large boulders, sharp corners), and used the software to spatially align the image with the shape model by minimizing the distance between corresponding reference points. Using a set of 20 reference points we achieved a root-mean square error (RMSE) between 3.1 and 9.9\,m (7.0\,m on average) for the alignment, depending on how many clearly defined landmarks are visible on the image that is to be aligned. Increasing the number of reference points did not further improve the RMSE. The point of view onto the projected image can then be changed in 3D to allow mapping from an optimum viewing angle.

Subsequently, we manually selected nodes along each feature of interest. The nodes each have three-dimensional XYZ point-coordinates in the comet-fixed Cheops reference frame. For ease of handling we exported the nodes as one polyline per feature. To aid visualization of the layering orientations, we determined best matching local plane solutions for each linear feature. The plane solutions consist of the normal vector ($\bm{n}$) and the reference point of the plane ($\bm{nb}$), which is the centroid of the nodes used to fit the plane. It also corresponds to the location of the base of $\bm{n}$. We found these plane solutions by applying a weighted least squares plane-fitting routine to the nodes of the feature \citep[Planefit,][]{schmidt_planefit_2012} in \textsc{matlab} \citep[release 2017a]{the_mathworks_inc._matlab_2017}). 

We assessed the uncertainty of the plane normal vectors $\bm{n}$ by means of a Monte Carlo simulation: The values of coordinates $X$, $Y$, and $Z$ of each node were varied, by a normally-distributed random value within the error of the nodes, around the measured coordinates $X_m$, $Y_m$, and $Z_m$. Those planes where $\bm{n}$ was poorly defined (variance of Monte Carlo results in at least one direction of $\geq$\,90\textdegree) were discarded from the pool of mapped features.

\subsection*{Ellipsoidal model fitting}

To evaluate whether linear features can be used as a standalone product to produce three-dimensional geological models, we made use of the same model defined by \citet{penasa_three_2017}. The model represents the layering of each lobe as a scalar field:
\begin{equation}
f(c_x,c_y,c_z) = k
\end{equation}
Function $f()$ is defined such that for any constant value of $k$ a single contour surface with ellipsoidal shape is determined, while the value of the scalar field represents a metric in the stratigraphic space ($0$ in the centre of the ellipsoidal model and increasing toward the outermost layers).

Function $f()$ is completely defined by eight parameters: $c_x, c_y$, and $c_z$ for the centre of the ellipsoidal model, $b$ and $c$ for the axial ratios and finally $\alpha, \beta$, and $\gamma$ for orienting the concentric ellipsoids in space. The parameters can be determined by maximizing the accordance of the model with the provided constraints. The orientation of the terraces, were used by \citet{penasa_three_2017} to provide observations of the gradient of the function $f()$ in a specific point $\bm{p}$, thus providing a modeling constraint of the type:
\begin{equation}
\label{eq:constraint_grad}
\nabla f(\boldsymbol{p})  = \boldsymbol{n} 
\end{equation}
where $\bm{n}$ is the normal to the surface element located in the point $\bm{p}$.

In this work we instead tested the use of polylines for modelling purposes. Each polyline is formed by segments which are expected to lie on a contour surface of the model and are thus tangential to the ellipsoidal shell passing for that location. Each segment can be defined by a pair of points $\bm{p}_1$ and $\bm{p}_2$ describing a direction in space, which can be represented as a unit vector:
\begin{equation}
\label{eq:segment}
\boldsymbol{\hat {t}} =\dfrac{\boldsymbol{p}_2 - \boldsymbol{p}_1}{\|\boldsymbol{p}_2 - \boldsymbol{p}_1\|} 
\end{equation}
where $\boldsymbol{\hat {t}}$ represents a constraint of tangent type \citep[e.g.][on this subject]{hillier_2014_threedimensional}:
\begin{equation}
\label{eq:constraint}
\left\langle\nabla f(\boldsymbol{p}), \boldsymbol{t} \right\rangle  = 0
\end{equation}
where $\bm{p}$ is the location of the observation (i.e. a reference point for the location of the segment). From these observations an angular misfit of the observation in respect to any model can be obtained by computing:
\begin{equation}
\label{eq:angle}
\vartheta =  \arccos\left(\left\Vert\dfrac{\nabla f(\boldsymbol{p})}{\| \nabla f(\boldsymbol{p}) \|} \times \boldsymbol{\hat t}\right\Vert\right)
\end{equation}

By minimizing the squares of the angles provided by \autoref{eq:angle} for each segment composing the mapped polylines, we were able to determine the most-likely parameters for the ellipsoidal model approximating the observations in this work. We employed a weighting strategy to ensure that each polyline contributes equally to the obtained solution. For this we divided each squared residual by the total number of segments of the specific polyline. Finally, we used a bootstrap strategy, based on the resampling of the polylines, to estimate the standard errors associated with each parameter.

\section{Results}

We used the PLW software to align a total of 34 OSIRIS images (\textbf{Table A2} in the supplementary material), covering most of the nucleus surface. On these images we mapped 171 linear features, of which 31 are terrace margins and 140 are strata heads. The mapped features are distributed approximately evenly between 67P's big and small lobe. The features extend laterally for 863\,m on average, ranging from 185 to 1925\,m. The feature length is calculated from the cumulative length of all segments connecting the polyline nodes. Most features contain between 9 and 20 segments (14 on average), and the average segment length is 38\,m. 

The uncertainty of each node results cumulatively from the error of the shape model ($\pm$ 0.3\,m), the resolution of the images (ca. 0.2\,\,m/px on average), and the image-alignment uncertainty in the PLW software (7\,m on average). Considering that the image resolution, and the error introduced by the alignment procedure, vary depending on the observation geometry and the distance from the camera, the overall uncertainty of each mapped node cannot be precisely quantified. However, based on the aforementioned considerations, it might be expected to be <\,10\,m.

The Monte Carlo error analysis showed that affecting the node coordinates $X$, $Y$, and $Z$ by random amounts between zero and 10\,m results in a variance of the normals $\bm{n}$ by less than 10\textdegree{} for most features (\textbf{Figure A1} in the supplementary material). For a small number of features, whose polylines have a low curvature, $\bm{n}$ is more strongly affected. As expected in such cases, the uncertainties show significant directional asymmetry and have large amplitudes only perpendicular to the main extension of the lineament. As \textbf{Figure A1(B)} shows, these cases are the exception in our data, and the node uncertainty does not have a substantial effect on the bulk of the layering orientations we reconstructed from the linear features.

\begin{figure}
	\includegraphics[width=\columnwidth]{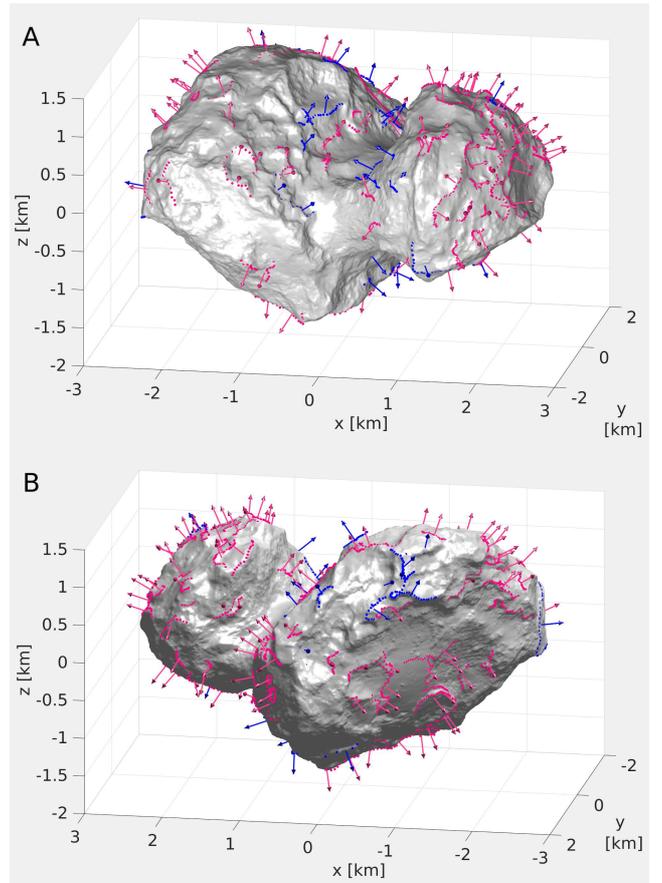}
    \caption{The SHAP7 shape model of comet 67P, showing the feature normals $\bm{n}$ at the bases $\bm{nb}$ for all mapped terrace margins (blue arrows) and strata heads (pink arrows). The dotted lines represent the mapped nodes along each feature. Coordinates are in the Cheops reference frame. \textbf{A:} `Front' view towards positive y-values; \textbf{B:} `Back' view rotated around the z-axis by 180\textdegree. Full-resolution images are available in the supplementary materials.} 
    \label{fig:figure2}
\end{figure}

An exemplary excerpt of the parameters of the plane solutions to the mapped features are listed in \autoref{tab:table1} (the complete \textbf{Table A1} is available in the supplementary material). The 3D orientation of vectors $\bm{n}$ is illustrated in \autoref{fig:figure2}. Qualitatively, by visual impression the normals are oriented perpendicular to the nucleus surface. For the big and small lobe separately, the normals are pointing outward from the respective lobe's gravitational centre.

\begin{table}
	\centering
	\caption{Exemplary excerpt of the normal vectors $\bm{n}$ and their bases $\bm{nb}$ to the plane solutions of the mapped strata heads on the big lobe. r is the radial distance of each vector base from the gravitational centre of the lobe. Values are in [km] in the comet-fixed Cheops reference frame. The complete tables are available in the supplementary material.}
	\label{tab:table1}
	\begin{tabular}{rrrrrrr}
		\hline
		$\bm{nb_x}$ & $\bm{nb_y}$ & $\bm{nb_z}$ & $\bm{n_x}$ & $\bm{n_y}$ & $\bm{n_z}$ & r\\
		\hline
    	0.316 & 0.695 & 0.616 & 0.632 & -0.425 & 0.648 & 1.3 \\
		-0.311 & 0.740 & 0.696 & 0.840 & -0.117 & 0.530 & 1.4 \\
		0.473 & 0.981 & -1.320 & 0.013 & 0.131 & -0.991 & 1.9 \\
		0.489 & 0.536 & -1.257 & 0.919 & 0.025 & -0.393 & 1.7 \\
		0.290 & -0.074 & -1.237 & 0.776 & -0.084 & -0.625 & 1.5 \\
		... & ... & ... & ... & ... & ... & ... \\
		\hline
	\end{tabular}
\end{table} 

\begin{figure}
	\includegraphics[width=\columnwidth]{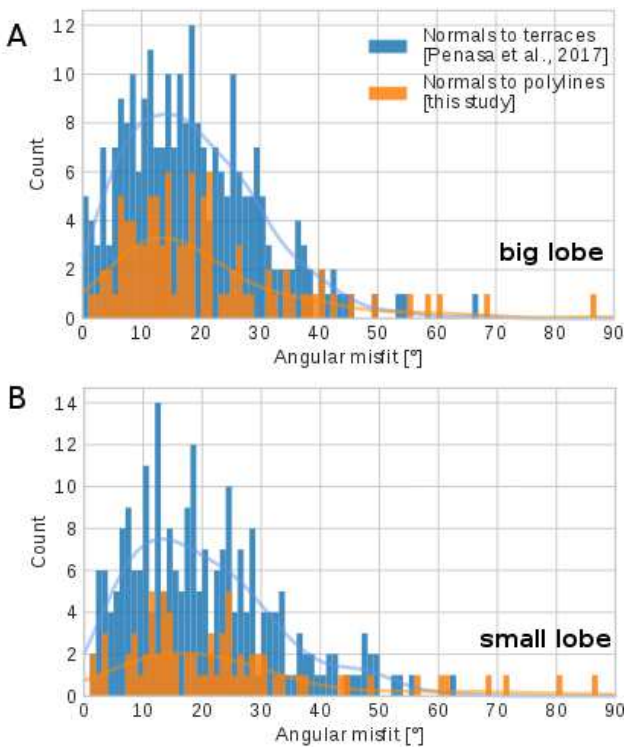}
    \caption{Histogram of the angular misfit between normal vectors $\bm{n}$ to the surface features and corresponding normal vectors $\bm{n_{ell}}$ of the ellipsoid model. The misfit for our linear features is plotted in orange, the misfit for the terraces in \citet{penasa_three_2017} is plotted in blue. \textbf{A:} Features on 67P's big lobe, \textbf{B:} Features on the small lobe.}
    \label{fig:figure3}
\end{figure} 

For the purpose of comparison, we fitted our own set of ellipsoids to the polylines of our linear features. The parameters we obtained for the best-fitting ellipsoidal model are summarized in \autoref{tab:table2}, next to the parameters achieved by \citet{penasa_three_2017}. The parameters are consistent for the big lobe within the achieved uncertainties, but we found a notable a misfit of the results for the small lobe. Our small lobe model is offset by 0.32\,km from their model (which amounts to ca. 40\% of the lobe's semi-minor axis, according to \citet{jorda_global_2016}), there is a minor difference in the ellipsoidal axis ratio, and a major mismatch in the rotational angles.

Finally, we compared the orientation of the plane normals $\bm{n}$ to the orientation of the ellipsoid surface for both \citet{penasa_three_2017} and our model. \autoref{fig:figure3} shows the angular misfit between $\bm{n}$ and the corresponding normal to the ellipsoid surface $\bm{n_{ell}}$, both at location $\bm{nb}$. Again, we find an overall low angular misfit for the big lobe (median 16.7\textdegree{}, panel A), and a larger misfit for the small lobe (median 19.6\textdegree{} with larger percentiles, panel B), indicating that the linear features are less congruent with the ellipsoid on the small lobe. \citet{penasa_three_2017} did not observe this dichotomy.

\begin{table*}
	\centering
	\caption{Maximum likelihood estimates of the ellipsoidal model parameters and relative 2$\sigma$ errors, modelled to our data and compared to \citet{penasa_three_2017}. Distances in km in the Cheops reference frame, and Tait-Bryan angles in degrees. BL means 67P's big lobe, SL is the small lobe; $b$ and $c$ are the axial ratios with respect to the a-axis ($a$\,=\,1).Values for which this study differs from \citet{penasa_three_2017} in excess of the uncertainties are highlighted in bold in the table.}
	\label{tab:table2}
	\begin{tabular}{lrrrrrrrrr} 
		\hline
		Parameter & \multicolumn{4}{l}{This study (171 linear features)} & & \multicolumn{4}{l}{\citet{penasa_three_2017} (483 terraces)} \\
                &  \  BL \ & \ 2$\sigma$ \ & \ SL \ & \ 2$\sigma$ \ & 
                &  \  BL \ & \ 2$\sigma$ \ & \ SL \ & \ 2$\sigma$ \ \\
		\hline 
  		$c_x$   & \ -0.55 \ & \ 0.12 \ & \ \textbf{1.35} \ & \ 0.09 \ & & \ -0.47 \ & \ 0.08 \ & \ 1.06 \ & \ 0.13 \ \\
        $c_y$   & \ 0.20 \ & \ 0.13 \ & \ -0.40 \ & \ 0.10 \ & & \ 0.32 \ & \ 0.08 \ & \ -0.35 \ & \ 0.07 \ \\
        $c_z$   & \ -0.15 \ & \ 0.11 \ & \textbf{0.13} \ & \ 0.08 \ & & \ -0.17 \ & \ 0.07 \ & \ 0.01 \ & \ 0.06 \ \\[1ex]
        b 	    & \ 0.80 \ & \ 0.08 \ & \ \textbf{0.85} \ & \ 0.08 \ & & \ 0.81 \ & \ 0.04 \ & \ 0.76 \ & \ 0.07 \ \\
        c  	    & \ 0.48 \ & \ 0.04 \ & \ 0.71 \ & \ 0.08 \ & & \ 0.55 \ & \ 0.03 \ & \ 0.70 \ & \ 0.07 \ \\[1ex]
        $\alpha$& \ 47.6 \ & \ 6.5 \ & \ \textbf{55.2} \ & 9.3 \ & & \ 44.8 \ & \ 4.3 \ & \ 28.1 \ & \ 9.3 \ \\
        $\beta$ & \ 7.3 \ & \ 10.7 \ & \ 4.2 \ & 18.0 \ & & \ 15.0 \ & \ 6.7 \ & \ -11.2 \ & \ 5.7 \ \\
        $\gamma$& \ 63.2 \ & \ 4.1 \ & \ \textbf{71.5} \ & 13.8 \ & & \ 66.3 \ & \ 3.9 \ & \ -7.3 \ & \ 34.4 \ \\
        \hline
        
	\end{tabular}
\end{table*} 

\section{Discussion and Conclusions}

While the layering-related linear features are not affected by sedimentation, we cannot rule out that the subset of terrace margins might be influenced by erosion and cliff collapse \citep[cf. e.g.][]{pajola_pristine_2017}. However, we took care to minimize this effect by accurately following the external border of the mapped edges, including any niches resulting from local breakoffs, thus preserving the orientation of the underlying layering.

The orientation of our feature normals $\bm{n}$  particularly in the `neck' region of the nucleus (\autoref{fig:figure2}), supports the widely accepted findings of previous works that the orientation of the layerings on the big and small lobe of 67P are independent from each other \citep[e.g.][]{massironi_two_2015, davidsson_primordial_2016}; it does not support a common envelope structure surrounding both lobes, nor the interpretation that both lobes represent fragments of a much larger, layered body \citep[as proposed by][]{rickman_comet_2015}.

We find that the internal structure of the nucleus, as deduced from the orientation of layerings mapped at the surface, is in agreement with the `onion-shell' model proposed by \citet{massironi_two_2015} and the concentric ellipsoidal shell model by \citet{penasa_three_2017}. Particularly for the big lobe, our results match those of \citet{penasa_three_2017} closely. We understand this as confirmation that the big lobe has a regular, concentric inner structure that follows a generally ellipsoidal configuration. Nevertheless, our data cannot confirm that the layerings are indeed connected into globally coherent shells. Approximating the minimal lobe circumferences as 6\,km and 8.6\,km, respectively, our measurements (including polylines of almost 2\,km length) might be mistaken to span a substantial portion of the nucleus surface. However, most of our polylines intentionally have a strong curvature and represent features with a continuous lateral extent of no more than a few hundred metres. This leaves room for the possibility of a discontinuously layered structure, as would be a consequence of e.g. the `talps' model \citep{belton_internal_2007} or layering by thermal processes \citep{belton_origin_2018}. Our findings would be also compatible with either concept.

For the small lobe, our results differ significantly from those of the other authors. Neither are the orientations of our proposed layerings clearly compatible with the ellipsoidal shell model based on its terraces, nor do the parameters of our own ellipsoidal model agree with those of the terrace-based model. This disagreement could either be explained by the circumstance that \citet{penasa_three_2017} mapped their terraces exclusively on the shape model, and thereby might have included some planar areas that are unrelated to the layerings. Another possible explanation is that the small lobe's inner structure has been affected by processes of evolution or deformation,
and thus cannot be unambiguously modelled by regular, concentric ellipsoids. In this case, a more complex model is required to represent the real geometry of the layerings.


\section*{Acknowledgements}

BKR was financially supported by the International Max Planck Research School (IMPRS) hosted by the Max Planck Institute for Solar System Research (G\"ottingen). This work benefited from insightful discussions with Matteo Massironi and his research group at the University of Padua. Geometrically calibrated images were made available by the OSIRIS team. The PLW software was provided by the Centre National d'\'Etudes Spatiales (CNES) in collaboration with Magellium. Borys Dabrowski is thanked for his help with the \textsc{matlab} coding.



\bibliographystyle{mnras}
\bibliography{bibliography}


\appendix
\section{Supplementary material}

Supplementary material is available online. It includes \textbf{Figures 1}, \textbf{2} and \textbf{3} in full resolution, and \textbf{Figure 2} as a \textsc{matlab} figure, and the complete table of feature normals (\textbf{Table A1}). \textbf{Table A2} lists all OSIRIS NAC images that were used for mapping, with the corresponding CODMAC calibration levels.

In addition, we included \textbf{Figure A1}, which is a visualization of the results of the Monte Carlo analysis of our plane normals, as described in the text. \textbf{A:} Plot of all Monte Carlo solutions. Each set of green arrows represents the spread of the normal vectors of a single linear feature. \textbf{B:} The Monte Carlo solutions of most features show low variability, as seen in this histogram of the standard deviations (2$\sigma$).

\bsp	
\label{lastpage}
\end{document}